\documentclass[9pt,twocolumn,twoside,pdflatex]{pnas-new}

\templatetype{pnasresearcharticle}

\setboolean{displaywatermark}{false}
\newcommand{\smallparallel}{{\mkern3mu\vphantom{\perp}\vrule depth 0pt\mkern2mu\vrule depth 0pt\mkern3mu}}
\usepackage{float}
\usepackage{physics}
\usepackage{color}
\usepackage{ulem}

\title{A direct link between active matter and sheared granular systems}

\author[a,1,*]{Peter K. Morse}
\author[b,1]{Sudeshna Roy} 
\author[c,1]{Elisabeth Agoritsas}
\author[b]{Ethan Stanifer}
\author[d]{Eric I. Corwin}
\author[b]{M. Lisa~Manning}

\affil[a]{Department of Chemistry, Duke University, Durham, North Carolina 27710, USA}
\affil[b]{Department of Physics, Syracuse University, Syracuse, New York 13244, USA }
\affil[c]{Institute of Physics, Ecole Polytechnique F{\'e}d{\'e}rale de Lausanne (EPFL), CH-1015 Lausanne, Switzerland}
\affil[c]{Department of Physics and Materials Science Institute, University of Oregon, Oregon 97403, USA}

\leadauthor{Morse} 

\significancestatement{There is not yet a robust theoretical framework predicting the dynamics of dense active matter, where energy is injected at the scale of constituent particles. Previous work has shown that some features of dense active matter are similar to those in dense disordered materials that are sheared globally from the boundaries. Using analytic and computational tools, we show that there is a direct correspondence between active matter and applied shear strain, which can in turn be used to help predict the behavior of dense active matter.}

\equalauthors{\textsuperscript{1}P.K.M., S.R., and E.A. contributed equally to this work}
\correspondingauthor{\textsuperscript{*}To whom correspondence should be addressed. E-mail: peter.k.morse@gmail.com}


\begin{abstract}
The similarity in mechanical properties of dense active matter and sheared amorphous solids has been noted in recent years without a rigorous examination of the underlying mechanism.
We develop a mean-field model that predicts that their critical behavior -- as measured by their avalanche statistics -- should be equivalent in infinite dimensions, up to a rescaling factor that depends on the correlation length of the applied field.
We test these predictions in 2d using a new numerical protocol, termed `athermal quasi-static random displacement', and find that these mean-field predictions are surprisingly accurate in low dimensions.
We identify a general class of perturbations that smoothly interpolate between the uncorrelated localized forces that occur in the high-persistence limit of dense active matter, and system-spanning correlated displacements that occur under applied shear.
These results suggest a universal framework for predicting flow, deformation, and failure in active and sheared disordered materials.
\end{abstract}

\dates{This manuscript was compiled on \today}
\doi{\url{www.pnas.org/cgi/doi/10.1073/pnas.XXXXXXXXXX}}

\begin{document}

\maketitle
\thispagestyle{firststyle}
\ifthenelse{\boolean{shortarticle}}{\ifthenelse{\boolean{singlecolumn}}{\abscontentformatted}{\abscontent}}{}

\dropcap{T}he statistical physics of active matter -- where energy is injected at the smallest scale, that of the particles themselves -- is highly nontrivial, exhibiting new features such as giant number fluctuations and motility-induced phase separation~\cite{narayan_long-lived_2007, cates_motility-induced_2015}. While comprehensive theories have been developed for many of these phenomena at low and intermediate densities \cite{cates_motility-induced_2015,marchetti_hydrodynamics_2013}, the behavior of highly dense, glassy active matter remains more mysterious.  Recent work by Henkes and collaborators~\cite{henkes_active_2011,henkes_dense_2020} highlights the important role of the potential energy landscape in constraining and dictating the behavior of dense active matter, which is in some ways similar to the situation in glasses excited by thermal fluctuations. Nevertheless, work by Berthier and collaborators emphasizes important differences between the dynamics of thermal and active glasses~\cite{berthier_non-equilibrium_2013, berthier_how_2017}
within the glassy potential energy landscape. Therefore, the large body of work on thermally excited glasses can not be transferred immediately to active glasses, and so a predictive theory for the dynamics of dense active matter remains elusive.

Meanwhile, the dynamics of athermal sheared disordered materials, where energy is injected at the largest scale, globally from the boundaries, has been the subject of intense study for decades.  A recent breakthrough allows an exact analytic solution for the behavior of slowly sheared systems in infinite dimensions, where interactions are exactly mean-field~\cite{rainone_following_2015,biroli_breakdown_2016,rainone_following_2016,urbani_shear_2017,biroli_liu-nagel_2018,altieri_mean-field_2019,parisi_theory_2020}. These results qualitatively explain many features in sheared 2- and 3-dimensional glassy solids. Perhaps more interestingly, new work suggests that the dynamical mean-field equations in infinite dimensions have the same structure regardless if the driving forces are generated by global shear or active forces on each particle~\cite{berthier_two-time-scale_2000,berthier_non-equilibrium_2013,agoritsas_out--equilibrium_2019,agoritsas_out--equilibrium_2019-1}, as all such forcing can be represented by memory kernels with the same functional form.

There is also evidence of similarities between sheared and active glassy systems in 2- and 3-dimensional simulations; recent studies have noted that in granular systems the two forcing mechanisms yield similar critical behavior~\cite{liao_criticality_2018}, large density fluctuations~\cite{henkes_active_2011,fily_freezing_2014}, effective temperatures~\cite{nandi_effective_2018}, aging behavior~\cite{mandal_multiple_2020}, and Eshelby deformations~\cite{mandal_extreme_2020}.

What is missing in the low-dimensional scenarios is a unifying picture as developed in infinite dimensions; to develop such a picture, it is necessary to first examine how and where discrepancies between shear and random forces appear. For example, Liao and Xu~\cite{liao_criticality_2018} noted that self-propelled particles driven by constant forces with the same magnitude in random directions will have the same diverging viscosity as their sheared counterparts~\cite{olsson_critical_2007,ikeda_disentangling_2013,bi_density-independent_2015} when jamming is approached, albeit with different critical exponents. Moreover, the values of the exponents can be changed by altering features of the forces on the self-propelled particles. Therefore, one wonders whether there may be a family of forcing fields, including shear and different types of self-propulsion, where all the resulting dynamics could be understood and predicted as part of a universal description of failure in jammed solids.

One hint about how such a framework might be constructed comes from the density of states that describes the spectrum of vibrational modes about a mechanically stable state in the potential energy landscape. More specifically, in low dimensions, it has been shown that the linear response of particles to either random forces in the limit of low rotational noise or long persistence length~\cite{henkes_active_2011,bi_motility-driven_2016,henkes_dense_2020}, or to shear~\cite{merkel_geometrically_2018} is dominated by the lowest eigenmode. Very close to an instability, this lowest eigenmode specifies the direction in the energy landscape with the lowest energy barrier~\cite{xu_anharmonic_2010}, and highlights the direction in which particles must move to leave one mechanically stable state and find another~\cite{malandro_relationships_1999,maloney_amorphous_2006}.

Taken together, these previous results suggest that in 2d and 3d materials there is a direct connection between how a disordered system traverses the energy landscape under shear and under random forces in the limit of zero rotational noise. Here we develop an exact infinite-dimensional mean-field theory prediction for the mechanical response of materials under shear and such active forces. We explicitly test this prediction by analyzing numerical simulations of soft spheres in two dimensions, and comparing dynamics under athermal quasi-static shear (AQS)~\cite{maloney_amorphous_2006} and a new constrained dynamics we term athermal quasi-static random displacements (AQRD). 

One goal of this manuscript is to establish AQRD as an interesting and important limit of active matter dynamics. In AQRD, each particle is displaced continuously along its own self-propelled direction. Typical active matter simulations study overdamped self-propelled particles that move under constant force, or equivalently constant velocity when the damping is homogeneous~\cite{marchetti_hydrodynamics_2013,cates_motility-induced_2015}. The direction of self-propulsion changes on a timescale called the persistence time, which is parameterized by the rotational noise. Therefore, AQRD is similar to self-propelled particles in the limit where the rotational noise is zero and the self-propelled velocity is slower than any other relaxation process inside the material.

An important difference between the two is, however, that active particles move under constant force, whereas AQRD particles move at constant displacement. This is in direct analogy to two different kinds of rheology experiments: (1) those where a system is subject to a constant shear force at the boundary, called "creep" experiments, and (2) those where the material is subject to a constant velocity condition at the boundary, called "constant strain rate" experiments. AQS is the zero-strain rate limit of the latter.
In this work, we focus on AQRD because simulations and experiments which control strain rate (or displacements) are known to be very useful for characterizing material properties, and so there is a large amount of data in the literature for comparison. We focus on the pre-yielding regime, corresponding with the "start-up" phase of a simulation or experiment where  the response depends strongly on the initial preparation of the material and the infinite-dimensional mean-field equations are solvable~\cite{rainone_following_2015,biroli_breakdown_2016,rainone_following_2016,urbani_shear_2017,biroli_liu-nagel_2018,altieri_mean-field_2019,parisi_theory_2020,agoritsas_mean-field_2021}. In contrast, stress-controlled creep experiments are fundamentally limited because the system can only cross energy barriers which are surmountable by the fixed applied stress, and under slow driving they exhibit complicated discontinuous stick-slip dynamics~\cite{dahmen_simple_2011,hayman_granular_2011}. Therefore, while our primary focus in this manuscript is on AQRD dynamics, we also introduce and study Athermal Quasistatic Random Force (AQRF) simulations, which are the random equivalent to creep experiments, and demonstrate that AQRF and AQRD are equivalent in linear response.

We next proceed to show that under shear (AQS) and random displacements (AQRD), scaling relations describing the avalanche statistics and the sampling of saddle points are identical and consistent with mean-field predictions, although the prefactors differ.  We hypothesize that differences in those prefactors, including the shear modulus, are governed by the correlation lengthscale associated with the imposed displacement field; in shear this length is the size of the box, while for completely random fields it is the size of individual particles. In addition, the mean-field calculation predicts that these prefactors are precisely determined by the distribution of the \textit{imposed} displacement field, which in turn causes fluctuations in strain between nearby particles.

Therefore, we systematically vary this correlation length in our simulations and find that the coefficients exhibit a systematic power-law scaling that matches mean-field predictions.  We also study the effect of material preparation on these results, demonstrating that shear and random displacement fields are similar even in ultrastable glasses.

Taken together, this demonstrates that shear can be considered as a highly-correlated special case of more general random displacements, and establishes AQRD as useful and interesting limit of active matter with a direct link to sheared systems.

\begin{figure}[!htb]
\centering
\includegraphics[width=\linewidth]{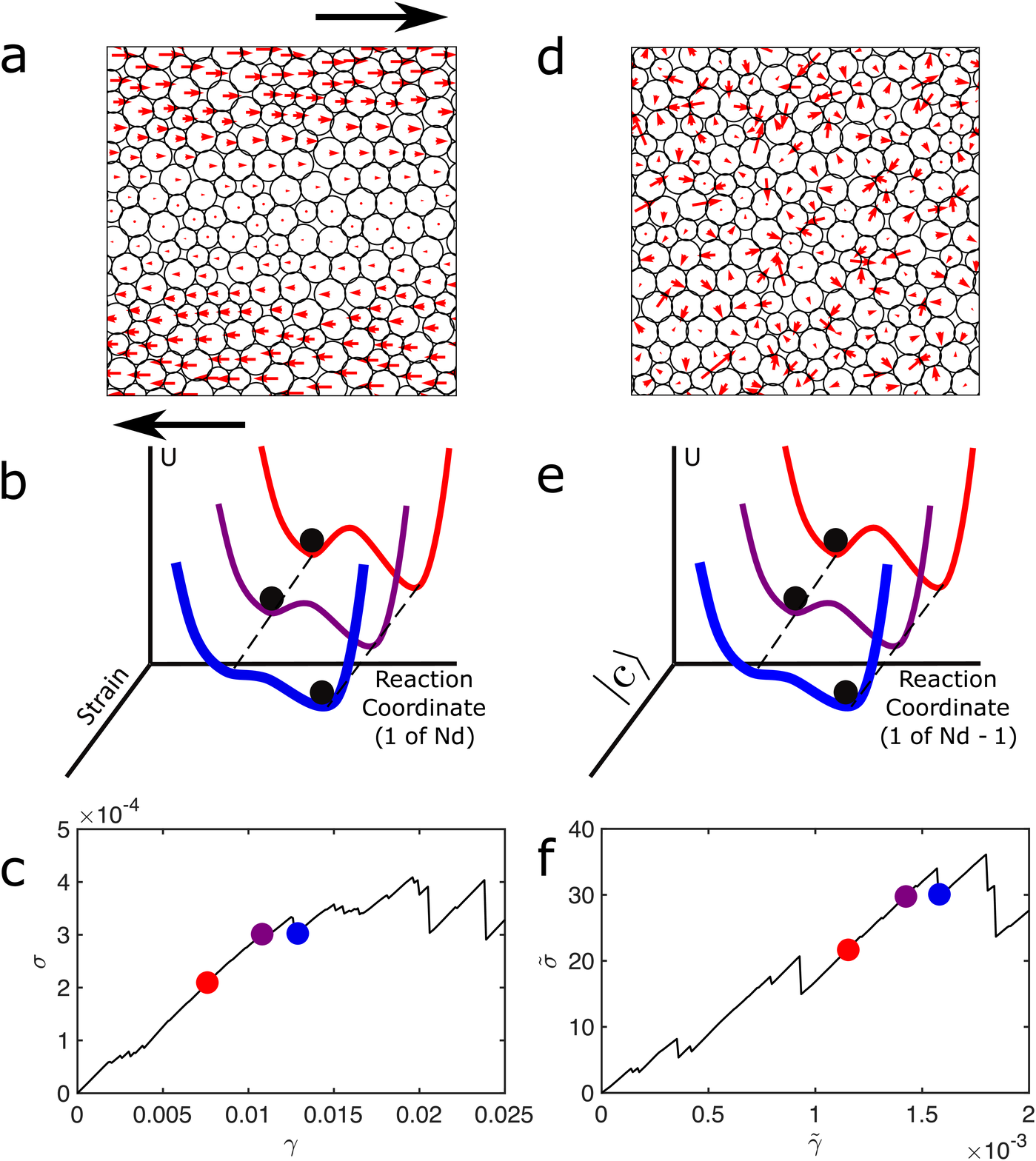}
\caption{\textbf{Two methods of traversing the energy landscape: AQS and AQRD}. \textbf{a)}~Forces applied to particles in an AQS ensemble.
\textbf{b)}~Potential energy landscape splitting out the ${Nd +1}$ degrees of freedom into $Nd$ (one of which is the reaction coordinate shown) and strain.
\textbf{c)}~Stress-strain curve showing that stress drop occurs when a saddle point in the reaction coordinate is reached by traversing along the strain coordinate.
\textbf{d)}~Forces applied to particles in a sample AQRD ensemble.
\textbf{e)}~Potential energy landscape splitting out the $Nd$ degrees of freedom (for fixed box shape) into ${Nd -1}$ (one of which is the reaction coordinate shown) and the vector along which random displacements are applied. \textbf{f)}~Random-stress vs. random-strain curve showing that random-stress drop occurs when a saddle point in the reaction coordinate is reached by traversing along the $\ket{c}$ coordinate. Highlighted points in (c) and (f) correspond with the curve of matching color in (b) and (e) respectively.}
\label{fig:saddlePoint}
\end{figure}

\section{Methods}

{\em Two ways of traversing the energy landscape---}
When constructing the energy landscape of allowed configurations, there are two types of variables that play \textit{a priori} different roles: \textit{state variables} are explicitly specified by the experimental or simulation protocol, while \textit{reaction coordinates} are free to vary under constraints imposed by state variables. For instance, a standard infinite temperature quench~\cite{ohern_jamming_2003} considers shear-strain to be a state variable during preparation, while shear-stabilization methods~\cite{dagois-bohy_soft-sphere_2012} treat strain as a reaction coordinate during preparation, regardless of how the strain variable is used afterwards. Therefore, the use of strain or the box degrees of freedom as state variables is merely an artifact of the way in which experiments or simulations are performed. Moreover, during an athermal \emph{quasistatic} perturbation, we adjust a state variable and then re-minimize the system by allowing all reaction coordinates to find their nearest local energy minima.

An applied shear strain, illustrated by the red arrows in Fig.~\ref{fig:saddlePoint}a, perturbs the system in its potential energy landscape. One way to represent this perturbation is to view the landscape as a function of the $Nd$ reaction coordinates (particle positions), so that adjusting the state variable (the magnitude of strain under simple shear) contorts the landscape in that $Nd$-dimensional space~\cite{malandro_relationships_1999,maloney_amorphous_2006, dagois-bohy_soft-sphere_2012}. As a system is sheared towards a saddle point, a nearby energy barrier is lowered until the system reaches the saddle point and moves downhill towards a new minimum.

It is equivalent to describe this process instead as moving in an ${Nd + 1}$ dimensional landscape where we explicitly push the system along the box degree of freedom, \textit{i.e.}~we control the state variable corresponding to the magnitude of simple shear strain, as shown in Figure \ref{fig:saddlePoint}b. In this framework, there are two types of saddle points: those parallel to the strain state variable, and those perpendicular to it. The ones perpendicular to the strain are the same as the saddles in the $Nd$-dimensional representation, whereas the saddles parallel to the strain correspond to the shear modulus changing sign, which does not correspond to an instability in a strain-controlled measurement~\cite{morse_differences_2020}.  

A second type of possible perturbation is a random displacement field, where we choose a random direction in configuration space $\ket{c}$ and promote it to a controlled state variable. An example field $\ket{c}$ is illustrated by the red arrows in Fig.~\ref{fig:saddlePoint}d.  Thus, after perturbing along $\ket{c}$, the system is free to relax along all directions perpendicular to $\ket{c}$, but motion along $\ket{c}$ is restricted via constrained minimization to the other $Nd-1$ dimensions. The saddles encountered in such dynamics are thus always perpendicular to $\ket{c}$, and we ask whether the distribution of saddles and their corresponding stress drops follow the same distribution as those encountered under shear strain.

{\em Numerical Model Description---}
We simulate $N$ Hertzian spheres in $d=2$ dimensions where $N$ is the number of particles. Except where specified when using ultrastable glasses, our systems are a 50-50 mixture of bidisperse disks with diameter ratio 1:1.4 to avoid crystalization. For the pressure sweep data, we prepare our systems at a target pressure by performing a standard infinite temperature quench~\cite{ohern_jamming_2003}, followed by FIRE minimization~\cite{bitzek_structural_2006} at a packing fraction such that we stay above the target pressure, followed by a careful decompression~\cite{morse_geometric_2016,morse_differences_2020}. For the correlation length sweep, we prepare our systems at a pressure of ${p = 0.0236 \pm 0.0004}$ via simple infinite temperature quench at a packing fraction $\phi=0.94$ \cite{ohern_jamming_2003}. In each case, we use the Hertzian contact potential
\begin{equation}
U = \frac{1}{5/2} \sum_{ij} \Theta(\varepsilon_{ij})\varepsilon_{ij}^{5/2}
\end{equation}
where $\Theta$ is the Heaviside function, $\varepsilon_{ij} = 1 - r_{ij}/(\rho_i + \rho_j)$ is the dimensionless overlap, $\rho_i$ is the radius of particle $i$, and $r_{ij}$ is the distance between particles $i$ and $j$.
All length scales are reported in natural units of the minimum particle diameter.

{\em Athermal Quasi-static Shear---}
Under the now-standard method of Athermal Quasi-static Shear (AQS)~\cite{maloney_amorphous_2006}, our system of particles is subject to simple shear via  Lees-Edwards boundary conditions where the periodic replicas in the $y$-direction are shifted by an amount $\gamma L_y$ in the $x$-direction, and $\gamma$ is the magnitude of simple shear which is the only non-zero entry in the strain tensor. After each small step in the applied strain ($\Delta \gamma = 10^{-4}$), a FIRE minimization algorithm~\cite{bitzek_structural_2006} is used to minimize the energy subject to the constraint that the box shape is held fixed (ensuring, therefore, that the strain tensor is defined by a single scalar, the shear strain). Therefore, AQS is equivalent to dynamics in the limit of zero strain rate -- where the material is sheared more slowly than any process or relaxation rate inside the material.

To facilitate comparison with the AQRD protocol described in the next section we emphasize that, in linear response and neglecting the effect of particle-particle interactions, shearing the boundary a distance $\gamma L_y$ along the $x$-direction is equivalent to displacing particles in the $x$-direction with a magnitude determined via the height of the system as given by ${u_i^\alpha=\gamma\delta^{\alpha x} (y_{i}-L_y/2)}$. Here $y_i$ is the $y$-coordinate of particle $i$, $L_y$ is the length of the box in the $y$-direction, and $\delta$ is the Kronecker delta function of $x$ and dimensional index $\alpha$ ~\cite{maloney_amorphous_2006}. An example of such a displacement field is shown in Fig.~\ref{fig:saddlePoint}a. The overall magnitude of this displacement vector field generated by an applied strain $\gamma$ is then given by ${|u(\gamma)|=\gamma\big[\sum_i \left(y_{i}-L_y/2\right)^2\big]^{1/2}}$. If we assume a uniform distribution of $y$-coordinate values, as one expects in an amorphous sample, the average magnitude is $|u(\gamma)|\approx\gamma L_y\sqrt{N/12}$. Therefore, an applied shear strain of $\gamma$ is equivalent to moving a distance $\gamma L_y\sqrt{N/12}$ along a normalized vector field, independent of dimension.

{\em Athermal Quasi-static Random Displacements---}
Similar to AQS, the system is initialized into a mechanically stable state at the bottom of a potential energy well with energy $U$ and $Nd$-dimensional position vector $\ket{x^{min}}$. The system is then displaced along an $Nd$-dimensional unitless vector $\ket{c}$ with elements $c_i$ and $\braket{c}{c}=1$. We explore different methods for choosing $\ket{c}$ described below. First, we define the {\em random strain} $\tilde{\gamma}$ of a scalar displacement $\tilde{u}$ along the vector $\ket{c}$ as
\begin{equation}
\tilde{\gamma} = \frac{\tilde{u}}{ L_y\sqrt{\frac{N}{12}}}.
\label{eqn:strainDisp}
\end{equation}
This definition ensures that strains ($\gamma$) in AQS can be directly compared to random-force strains ($\tilde\gamma$) in AQRD, where both are unitless.   

Starting from positions $\ket{x^{min}}$ and displacing by an amount $\tilde{u}$, new positions are then ${\ket{x} = \ket{x^{min}} + \tilde{u}\ket{c}}$, but they are not in a local energy minimum with respect to the reaction coordinates. Therefore, we must evolve the system using a constrained minimization that imposes an external force $\ket{F^{ext}} = -\lambda \ket{c} $,  where $\lambda$ is the Lagrange multiplier, which prevents any motion along $\ket{c}$.

We calculate how such displacements induce changes to the internal stress of the system, in direct analogy to stress-strain curves for AQS. The stress induced by the field $\tilde{\gamma}$ is given by
\begin{equation}
\tilde\sigma = \frac{1}{A}\frac{d U}{d \tilde\gamma} = \frac{1}{A}\sum_{i=1}^N \bigg(\frac{\partial U}{\partial x_i^\smallparallel}\frac{dx_i^\smallparallel}{d\tilde\gamma} + \frac{\partial U}{\partial x_i^\perp}\frac{d x_i^\perp}{d \tilde\gamma}\bigg).
\end{equation}
where $A=L_xL_y$ is the area, and we have split the particle motion $\mathbf{x}_i$ into components which are parallel or perpendicular to $\mathbf{c}_i$ as $x_i^\smallparallel$ and $x_i^\perp$ respectively. By definition, $\frac{\partial U}{\partial x_i^\perp} = -F_i^\perp$ and since we minimize force with respect to the particle position, $F_i^\perp = 0$. Thus, the total residual force $\mathbf{F}_i$ on each particle $i$ is parallel to $\mathbf{c}_i$. Furthermore, we note that ${\frac{dx^\smallparallel_i}{d\tilde\gamma} = \mathbf{c}_iL_y\sqrt{\frac{N}{12}}}$, resulting in the definition of the {\em random stress}
\begin{equation}
\label{eq-def-random-stress}
\tilde\sigma = -\frac{1}{L_xL_y}\sum_{i=1}^N {\mathbf{F}_i\cdot\mathbf{c}_i} L_y\sqrt{\frac{N}{12}} = -\bra{F}\ket{c} \frac{1}{L_x}\sqrt{\frac{N}{12}}.
\end{equation}
This is a generalization of the derivation for shear stress in AQS developed Maloney and Lemaitre \cite{maloney_amorphous_2006}.
Throughout this manuscript, we use variables with a \emph{tilde} to denote observables that are the AQRD equivalent to AQS counterparts.

In practice, we evolve the system by taking steps of $10^{-4}$ in the random strain $\tilde{\gamma}$, and after each step we use FIRE minimization \cite{bitzek_structural_2006} to find the constrained local minimum.
Thus, instead of applying forces in the FIRE-calculated gradient direction $\ket{F}$, we apply them along $\ket{F} - \bra{c}\ket{F}\ket{c}$. We impose a stopping condition when every component of the total excess force on every particle is less than a cutoff value of $10^{-14}$, set to ensure particle positions to double precision. By construction, there is no drift velocity in the system.

\begin{figure*}[!htb]
\centering
\includegraphics[width=0.9\linewidth]{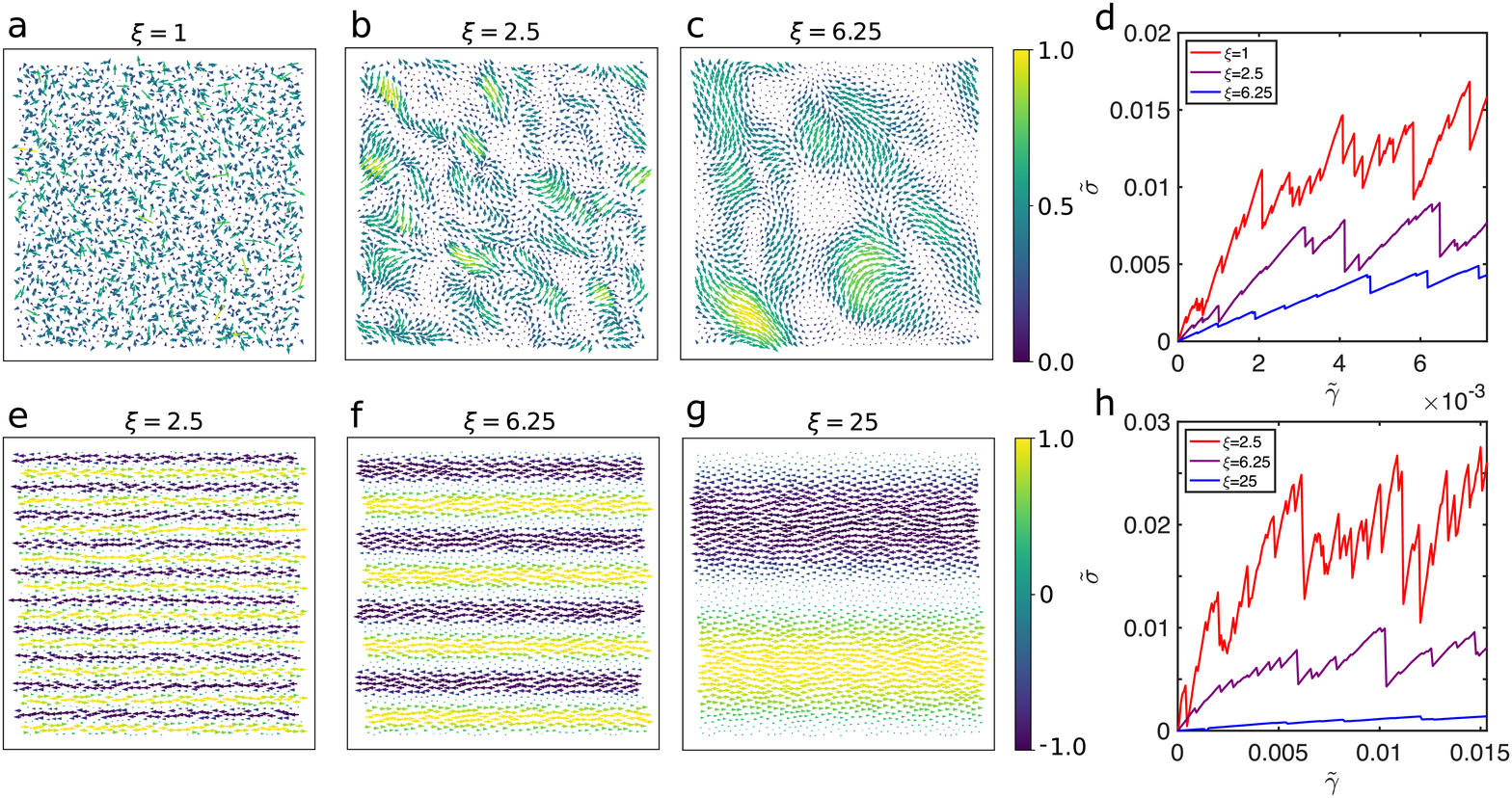}
\caption{{\bf Effect of random field correlation length $\xi$ on the mechanical response.} {\bf a-c} Snapshots of Gaussian correlated fields (GCF) with correlation lengths \textbf{a)}~$\xi = 1$, \textbf{b)}~$\xi = 2.5$, \textbf{c)}~$\xi = 6.25$. \textbf{d)}~Example random-stress vs. random-strain curves for random fields with different correlation lengths. \textbf{e-g} Snapshots of  wave-like correlated fields (WCF) with wave lengths \textbf{e)}~$\xi = 2.5$, \textbf{f)} $\xi = 6.25$, \textbf{g)} $\xi = 25$. \textbf{h)} Example random-stress vs. random-strain curves for wave-like fields with different correlation lengths. In all graphs, we use $N = 2048$, $\phi=0.94$, and thus $L_x=L_y=50.3$.}
\label{fig:stress_strain}
\end{figure*}

We generate the fields $\ket{c}$ for AQRD using two different methods: one based on random Gaussian fields and another based on plane waves. The Gaussian random fields, which are spatially correlated over a characteristic length scale $\xi$, are generated using a standard Fourier transform method that respects the periodic boundary conditions. A detailed description is given in the supplement.
Fig.~\ref{fig:stress_strain}a-c illustrates the random vector $\ket{c}$ generated from the correlated Gaussian random field for different correlation length $\xi = 1$, $2.5$ and $6.25$, respectively. To test whether features we observe are dependent only on the correlation length, or whether other features of the field structure are important, we also generate plane-wave-like fields
where the $x-$components of the vectors are a sine function of the $y-$coordinate of the particle positions, and the $y-$components of the vectors vanish. For such fields, we define the correlation length scale to be half the chosen wavelength. Fig.~\ref{fig:stress_strain}e-g illustrates the random vector $\ket{c}$ generated in wave-like pattern for different correlation lengths $\xi = 2.5$, $6.25$ and $25$ respectively. The corresponding displacement field associated with shear under Lees-Edwards boundary conditions is equivalent to a plane wave with a wavelength $2L_y$, which is clear from Fig.~\ref{fig:saddlePoint}a.

While this version of AQRD applies displacements in a direct analogy to a strain-controlled experiment, we also study a stress-controlled version of random forcing, denoted athermal quasistatic random forcing (AQRF), which is an exact limit of standard active matter simulations. Details can be found in the supplement. Fig~\ref{fig:AQRF}a-c compares the dynamics under AQRD (black) and AQRF (red) for a system with the same initial conditions. In linear response (\textit{i.e.}~until the first stress drop in AQRD), the two curves are exactly equivalent. More broadly, until the macroscopic yielding transition (at about 6 \% strain), stress drops in AQRD are often associated with slip events in AQRF, and the curves still largely follow each other, similar to results in sheared particle systems. Fig~\ref{fig:AQRF}d demonstrates that these similarities persist over a large ensemble.  Together, these data indicate that AQRD and AQRF sample similar features of the potential landscape in the pre-yielding regime, which is also consistent with a full derivation of the mean-field theory~\cite{agoritsas_mean-field_2021}. This confirms that AQRD is a useful proxy for active matter simulations in the limit of zero rotational noise, and so we focus on AQRD in what follows.

\begin{figure*}[!htb]
\centering
\includegraphics[width=\linewidth]{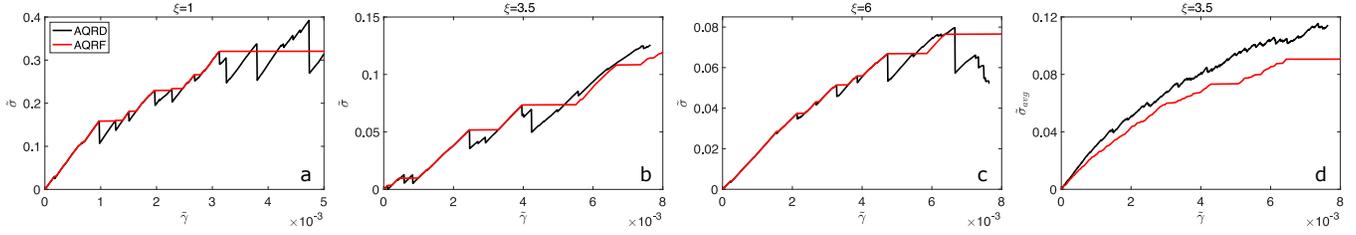}
\caption{\textbf{Comparison of AQRD and AQRF sampling mechanisms.} Three generic sample systems with $N=2048$ and $\phi=0.94$ are generated and then perturbed by a Gaussian correlated field (GCF) $\ket{c}$ with \textbf{a)} $\xi=1$, \textbf{b)} $\xi=3.5$, and \textbf{c)} $\xi=6$. The driving mechanism is varied between AQRD -- in direct analogy to a strain-controlled measurement -- and AQRF -- in direct analogy to a stress-controlled measurement. In linear response, the curves are exactly equivalent, but while AQRD systems experience stress drops, AQRF measurements are punctuated by slip events, wherein the system rearranges until it can support the applied stress. As such, in AQRF, the system does not sample local minima in the energy landscape. \textbf{d)} Nevertheless, the curves can be averaged (in this case, over 30 realizations) to give the bulk response. In the pre-yielding regime, we see hat the average response is the same, indicating that both mechanisms sample similar features of the energy landscape.}
\label{fig:AQRF}
\end{figure*}

\section{Results}
{\em Mean-field results---}%
The limit of infinite dimension provides an exact benchmark to investigate properties of structural glasses~\cite{parisi_theory_2020,charbonneau_fractal_2014},
and has been successfully used, for instance, to study quasistatic shear or compression~\cite{rainone_following_2015,biroli_breakdown_2016,rainone_following_2016,urbani_shear_2017,biroli_liu-nagel_2018,altieri_mean-field_2019,parisi_theory_2020}. In this framework, we can show that AQS and AQRD are strictly equivalent upon a simple rescaling of the accumulated strain, with a dependence on the correlation length $\xi$. The full derivation is provided in Ref. \cite{agoritsas_mean-field_2021}.

In order to implement a local strain vector $\ket{c} \in \mathbb{R}^{Nd}$ as in AQRD, we assign to each particle a random local strain ${{\bf c}_i}$ drawn from a Gaussian distribution with zero mean defined by:
\begin{equation}
\label{eq-def-individual-strains-distr}
\begin{split}
& \overline{{\bf c}_i}=0
\, , \quad
\overline{{\bf c}_i\cdot{\bf c}_j} = \Xi \, f_{\xi}(\vert {\bf r}_{ij} (0) \vert)/d \, ,
\\
& \text{with} \quad
f_{\xi}(x)=e^{-x^2/(2\xi^2)}/\sqrt{2 \pi \xi^2}
\, ,
\end{split}
\end{equation}
where the overline denotes the statistical average over the quenched random strain field,
$\Xi$ is a tunable amplitude which has the units of a length (so that the strains remain unitless), and ${{\bf r}_{ij} (t)}$ is the distance between particles $i$ and $j$ at time $t$ (and we focus here on the initial configuration).
For simplicity here we have assumed the fluctuations in the field can be described by a normalized Gaussian function
with a finite correlation length ${\xi>0}$.
However, we emphasise that this simplifying condition on ${f_{\xi}(x)}$ does not meaningfully affect the main results and the general case is treated in Ref.~\cite{agoritsas_mean-field_2021}. Finally we include an explicit scaling with dimension $d$ so that the fluctuations in $c$ scale with dimension in the same way as fluctuations in the local strain field in AQS.

In the infinite-dimensional limit, the complex many-body dynamics of pairwise interacting particles becomes exactly mean-field.
It can then be reduced to an effective scalar stochastic process for the fluctuating gap between particle pairs, ${h_{ij}(t) = d \left( \vert {\bf r}_{ij}(t) \vert /\ell - 1 \right)}$ where $\ell$ is the typical distance between particles and ${h_{ij} \sim \mathcal{O}(1)}$ \cite{maimbourg_solution_2016,szamel_simple_2017,agoritsas_out--equilibrium_2019-1,agoritsas_out--equilibrium_2019}. To compare the mean-field gap directly with the soft spheres in our simulations, we can use the relationship $h_{ij} = -d \varepsilon_{ij}\frac{\rho_i + \rho_j}{\ell}$.
The dynamics are then governed by the distribution of the relative strains ${ c_{ij} \equiv \norm{{\bf c}_i -{\bf c}_j}}$, which are uncorrelated in the limit ${d \to \infty}$ for distinct pairs of particles (consistent with the mean-field assumption).
The variance of a given pair $\overline{c^2_{ij}}$, however, still encodes the spatial correlation of individual local strains, through the following quantity:
\begin{equation}
\mathfrak{F}\left(\Xi, \ell, \xi \right)
= d \ell^2\, \overline{c_{ij}^2} =2 \ell^2 \Xi \left[ f_{\xi}(0) - f_{\xi}(\ell) \right] \, ,
\label{eqn:Fcorr}
\end{equation}
%
%
which can be straightforwardly computed for a given choice of $f_\xi$, or directly measured in numerical simulations.
By adapting the derivation of the mean-field description for shear presented in Ref.~\cite{agoritsas_out--equilibrium_2019}, we find that AQS and AQRD are strictly equivalent in infinite dimension, provided that we rescale the accumulated strain by a factor ${\sqrt{\mathfrak{F}}/\ell}$, so that it is directly controlled by the variance of relative strains ${\overline{ c_{ij}^2}}$.

For the quasistatic stress-strain curves and the elastic modulus, we specifically predict that the random strain $\tilde{\gamma}$ can be written in terms of the AQS shear strain $\gamma$, and therefore the random-displacement stress $\tilde{\sigma}$ and the random-displacement modulus $\tilde{\mu}$ can also be easily scaled:
\begin{equation}
\begin{split}
\gamma_\mathrm{MF} \equiv \tilde{\gamma}_\mathrm{MF} \frac{\sqrt{\mathfrak{F}}}{\ell}
 \: \Rightarrow \: \left\lbrace \begin{array}{l}
 \sigma_\mathrm{MF} = \frac{\ell}{\sqrt{\mathfrak{F}}} \, \tilde{\sigma}_\mathrm{MF} ,
 \,
 \\
  \mu_\mathrm{MF} =  \frac{\ell^2}{\mathfrak{F}} \tilde{\mu}_\mathrm{MF} ,
 \,  \end{array} \right.
\end{split}
\label{eqn:mfpred}
\end{equation}
where the MF subscripts emphasize that this is a mean-field prediction, whose validity should be tested in lower dimensions.

We emphasize that the infinite-dimensional calculation predicts that ${\mathfrak{F}/\ell^2}$ is thus the key quantity to make the AQRD random stress-random strain curves (and other such mean-field observables) collapse onto their AQS counterparts.
This quantity is solely prescribed by the statistical features of the \emph{input} field that we chose to consider.
Simply put, ${\mathfrak{F}/\ell^2}$ is the variance in the strain of the input field -- \textit{i.e.}~a measure of the distribution of relative strain between particles -- and it completely governs the dynamics of the system.

Under our assumption that ${f_{\xi}(x)}$ is a normalized Gaussian function as in ~\eqref{eq-def-individual-strains-distr},
we can straightforwardly compute  $\mathfrak{F}$ from ~\eqref{eqn:Fcorr}.
By Taylor-expanding $\mathfrak{F}$ in the limits ${\ell/\xi \ll 1}$ and ${\ell/\xi \gg 1}$, and keeping only the leading terms, we predict a crossover of the elastic modulus $\xi$-dependence depending on the ratio ${\ell/\xi}$, with ${\mathfrak{F} \sim 1/\xi}$ at ${\ell/\xi \gg 1}$ and ${\mathfrak{F} \sim 1/\xi^{3}}$ at ${\ell/\xi \ll 1}$ \cite{agoritsas_mean-field_2021}. The specific case of global applied shear strain corresponds to the latter case, as $\xi$ is of the order of the system size for shear. In both cases, this implies that the elastic modulus decreases with increasing $\xi$, as we will demonstrate numerically below.
This matches with physical intuition: it is less efficient to deform a glass with more correlated local strains, \textit{i.e.}~with a larger correlation length.
The most extreme case is to consider an infinite $\xi$: if all particles are driven with the same vector ${\bf c}_i$,
the whole system is simply translated in space and its effective strain is strictly zero, consistently with having no variance of relative strains (${\mathfrak{F}=0}$).
In particular, ~\eqref{eqn:mfpred} states that in mean-field, AQS is a special case of AQRD, with $\mathfrak{F}/\ell^2=1$.  
See the supplement for a scaling argument in \emph{finite} dimension supporting this mean-field picture.

{\em Numerical results for random stress vs. random strain ---}
We next test the mean-field prediction in numerical simulations in 2D. Our first observation is that AQS and AQRD give rise to qualitatively similar stress-strain and random stress-random strain curves, as highlighted in Figs.~\ref{fig:saddlePoint}c and ~\ref{fig:saddlePoint}f.  Elastic branches -- where the stress rises linearly with the strain -- are punctuated by points where the system crosses a saddle point instability, causing a stress drop and particle rearrangements as the system transitions to a new energy minimum. The magnitude of the stress drop quantifies the size of the rearrangement event.

\begin{figure}[t]
\centering
      \includegraphics[width=0.8\linewidth]{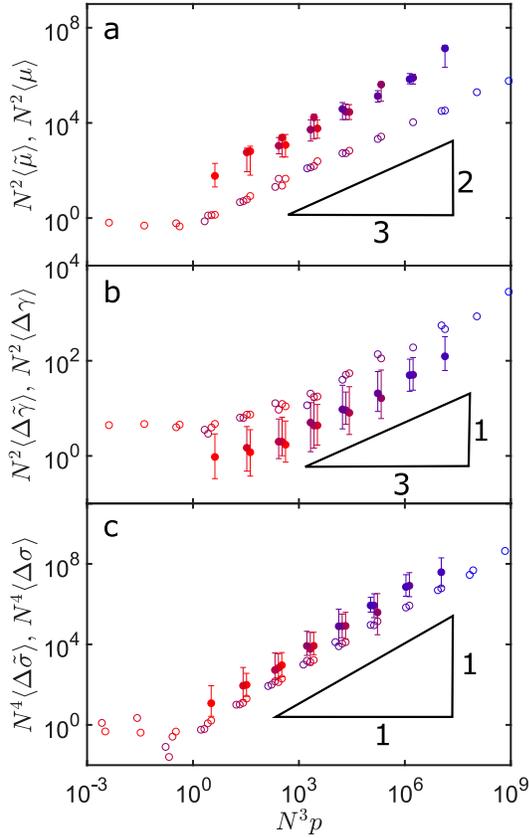}
  \caption{\textbf{System size and pressure dependence of landscape statistics.} \textbf{a)}~Local shear modulus $\mu$,
  \textbf{b)}~Strain distance between rearrangements $\Delta \tilde{\gamma}$, and \textbf{c)}~stress drops across rearrangements $\Delta \tilde{\sigma}$ as a function of $N^3p$ to show collapse with system size and pressure. AQRD with completely uncorrelated random fields is shown with closed circles, while AQS data is shown with open circles. Error bars represent the middle 60\% of the distribution and are only shown for AQRD for visual clarity, but are approximately the same for AQS. Colors represent system sizes $N=64$ (red), $128$, $256$, $512$, $1024$ (blue) in an even gradient. Corresponding pressures are $p = 10^{-2}$, $10^{-3}$, and $10^{-4}$.}
\label{fig:landscapeStatistics}
\end{figure}

In AQS, the stress averaged over many such stress drops gradually rises until about 6-7\% strain, at which point the systems yields. After the yielding point, the average stress remains constant as a function of strain. Moreover, the local shear modulus $\mu$, defined as the slope of the stress-strain curve along elastic branches, is significantly different from the macroscopic coarse-grained shear modulus $\mu_\mathrm{global}$, defined as the ratio of the average stress at yield to the average strain at yield. This observation is directly related to marginal stability~\cite{lin_mean-field_2016}, and can be qualitatively predicted from infinite dimensional analytic theory~\cite{rainone_following_2015,biroli_breakdown_2016,rainone_following_2016,urbani_shear_2017,biroli_liu-nagel_2018,altieri_mean-field_2019,parisi_theory_2020}.

To develop a more quantitative comparison between AQS and AQRD, as predicted in ~\eqref{eqn:mfpred}, we focus on three metrics that quantify how AQS and AQRD sample phase space in the pre-yielding regime:
\textit{(i)}~the distribution of local shear/random-displacement moduli $\mu$ and $\tilde{\mu}$ along elastic branches,
\textit{(ii)}~the distribution of (random) strain intervals $\Delta \gamma$ and $\Delta \tilde{\gamma}$ between stress drops,
and \textit{(iii)}~the distribution of (random) stress drop magnitudes $\Delta \sigma$ and $\Delta \tilde{\sigma}$. We use $\langle \Delta \tilde{\gamma} \rangle$ and $\langle \Delta \tilde{\sigma} \rangle$ to denote quantities which are explicitly averaged over all elastic branches in the pre-yielding regime.

{\em Scaling of observables with system size and pressure---} 
Previous work has analyzed these statistics in AQS as a function of system size $N$ and pressure $p$~\cite{shang_elastic_2020,morse_differences_2020,franz_mean-field_2017}, as such data helps constrain continuum so-called `elasto-plastic' models to predict features of avalanches in granular matter.  In addition, the size of a rearrangement provides interesting information about the nonlinear features of the potential energy landscape, as it is one way of quantifying how far the system has to travel from a saddle point to find a nearby local minimum.  The size of avalanches in AQS, quantified by the magnitude of the stress drops and other metrics, is known to exhibit power-law scaling with a large-scale cutoff, and the power law has different exponents on either side of the yielding transition~\cite{shang_elastic_2020}. In the pre-yielding regime the average stress drop is well defined, and changes in a systematic way with system size and pressure.  Previous work by some of us~\cite{morse_differences_2020} demonstrated that in AQS the average stress drop exhibits two regimes: a finite-size regime when $N^3p \ll 1$ in which the size of stress drops remains constant, and a second regime when $N^3p \gg 1$ where the stress drops scale as $\langle\Delta \sigma\rangle \sim \frac{p}{N}$, which is illustrated by the open symbols in Fig.~\ref{fig:landscapeStatistics}c.

\begin{figure*}[t]
\centering
\includegraphics[width=\linewidth]{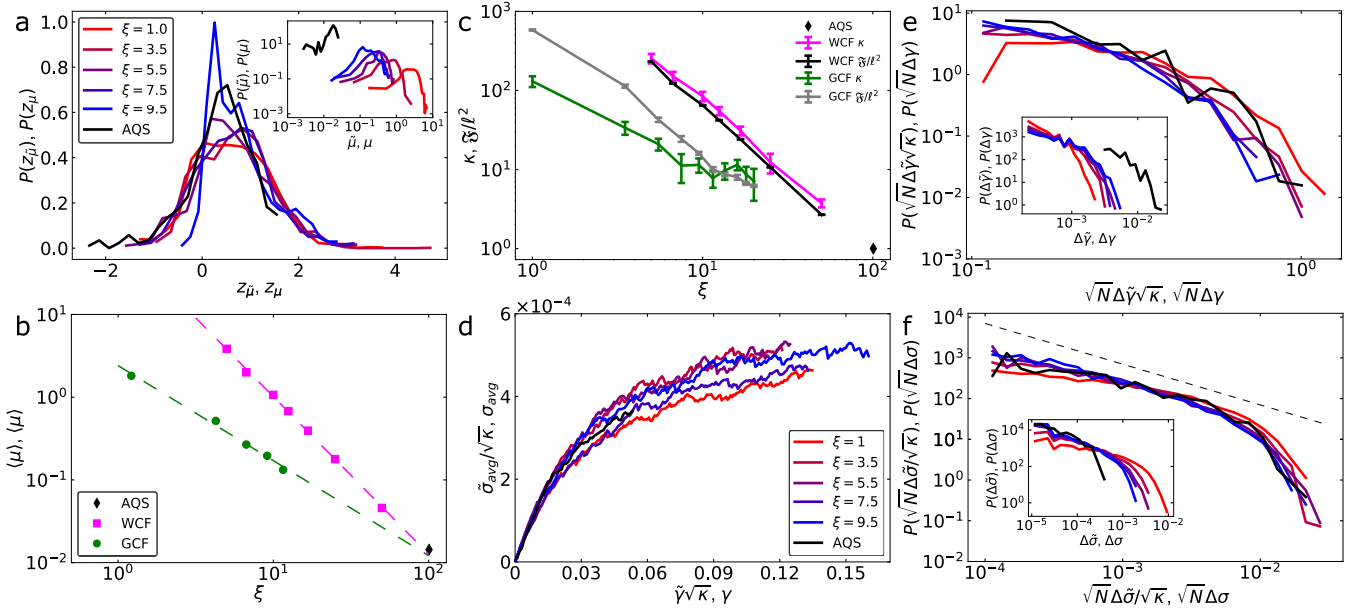}
\caption{{\bf Collapse of landscape statistics with correlation length.}
\textbf{a)}~Probability distribution of the local effective moduli $\tilde{\mu}$ and $\mu$ (inset) and the recentered $z_{\tilde{\mu}}$ and $z_{\mu}$ (rescaled by their standard deviations) in GCF systems with $\xi=1$ (red) through $\xi=9.5$ (blue) compared with $\mu$ of AQS (black). 
\textbf{b)}~The average effective modulus decreases as a function of correlation length in both WCF and GCF ensembles. All curves approach the AQS value (black diamond), and dashed lines are best fits for $\tilde{\mu}_{\text{WCF}}$ with slope $-1.9$ (magenta)  and  $\tilde{\mu}_{\text{GCF}}$ with slope $-1.1$ (green) respectively, consistent with the mean-field prediction of a slope between $-1$ and $-3$.
\textbf{c)}~A comparison of $\mathfrak{F}/\ell^2$ computed directly via the variance of the field $\ket{c}$ (black and gray lines) and the initial modulus ratio $\kappa=\tilde{\mu_0}/\mu_0$ (magenta and green lines).
\textbf{d)}~Collapse of average stress-strain curves for GCF random fields onto average AQS stress-strain curve using Eq.~\eqref{eqn:mfscale}. Here $\tilde{\sigma}_{avg}$ and $\sigma_{avg}$ denote an average over configurations but not all elastic branches. We are additionally able to collapse the distributions of \textbf{e)}~the effective strain interval $\sqrt{N}\Delta\tilde\gamma\sqrt{\kappa}$ and \textbf{f)}~the effective avalanche size $\sqrt{N}\Delta\tilde\sigma/\sqrt{\kappa}$, by appropriate scaling of the raw data (insets). Data shown is for GCF, with WCF shown in the supplement. Additionally, finite size scaling showing the empirical collapse with the given factors of $\sqrt{N}$ is shown in the supplement. The avalanche distribution agrees with the reported slope of $-1$ (dashed black line) \cite{shang_elastic_2020} given as a guide to the eye. }
  \label{fig:correlationCompare}
\end{figure*}

Therefore, we first study the statistics of stress drops for the simplest choice for the AQRD vector field $\ket{c}$ -- an uncorrelated random field (GCF with $\xi=1$), which is also most similar to typical self-propelled particle simulations for active matter. The closed symbols in Fig.~\ref{fig:landscapeStatistics}c correspond to stress drop statistics in the pre-yielding regime for an ensemble of 50 different initial configurations at each value of $N$ and $p$, showing that precisely the same scaling is seen in AQRD. This highlights that the zero-pressure limit of the avalanche statistics under AQRD is singular, just as in AQS. Although the scaling is identical there is clearly a shift in the prefactors, which we return to in the next section.

In addition to the magnitude of the stress drops, the strain between saddle points or rearrangements provides another window into the statistical features of the complex potential energy landscape.  Fig.~\ref{fig:landscapeStatistics}b clearly shows that the mean strain interval between rearrangements scales as $\langle\Delta\tilde{\gamma}\rangle \sim \frac{p^{1/3}}{N}$ in both AQS (open circles) and AQRD (closed circles). 
Additionally, we measure the average shear modulus between rearrangements, which scales  as $\langle\mu\rangle \sim p^{2/3}$ for both AQS and AQRD as shown in Fig.~\ref{fig:landscapeStatistics}a. 

{\em Effects of spatially correlated forcing---} Although the scaling exponents of the previous section are precisely the same under both AQS and AQRD dynamics, it is clear that there is a systematic offset in the prefactors, despite the fact that care was taken to ensure the definition of effective strain in each case is equivalent.

To understand the origin of this difference, we vary the correlation length $\xi$ of the normalized AQRD vector field $\ket{c}$ measured in units of the smaller particle diameter and use Gaussian correlated fields (GCF) and wave-like correlated fields (WCF), as described in the methods section and illustrated in Fig.~\ref{fig:stress_strain}. For these analyses, system size $N = 2048$ and packing fraction $\phi = 0.94$ are fixed and known to be far from the singular limit. These parameters produce ${L_x = L_y = \sqrt{\frac{N\pi(1+1.4^2)}{8\phi}} = 50.3}$, where $1$ and $1.4$ are the diameters of the two species of particles.

Examples of $\ket{c}$ for both GCF and WCF are shown in Fig.~\ref{fig:stress_strain}. In each case, the random stress vs. random strain curves exhibit qualitatively similar features, with elastic branches punctuated by stress drops. The overall magnitude of the stress scale changes dramatically, where larger stresses are associated with smaller correlation lengths.

In order to test the prediction of ~\eqref{eqn:mfpred}, we first investigate the statistics of the local shear modulus, $\tilde{\mu}$, shown for the GCF data in the inset to Fig.~\ref{fig:correlationCompare}a. The GCF distributions shifted by the mean $z_{\tilde{\mu}}$ and scaled by the standard deviation do not collapse as shown in the main panel Fig.~\ref{fig:correlationCompare}a. However, the average is well-defined for both GCF and WCF data sets, and decreases with increasing $\xi$ (Fig.~\ref{fig:correlationCompare}b). Specifically, both data sets are consistent with $\tilde{\mu}$ being a power law function of $\xi$.  

We note that the AQS data point shown by the black diamond falls on both of the lines describing GCF and WCF data, respectively.  This must be the case, as the only input field with correlation length equal to the box size that obeys the necessary constraints -- namely that the field has zero mean and respects the periodic boundary conditions  -- is the one corresponding to simple shear (See supplement 1B for more details). Nevertheless, this observation confirms that shear is a special case of a more generalized response to displacement fields.

Next, we define a new variable, $\kappa$, as the initial random-displacement modulus $\tilde{\mu}_0$ normalized by the initial shear modulus $\mu_0$: $\kappa \equiv \tilde{\mu}_0 / \mu_0$. We then explicitly test the mean-field prediction for the shear modulus, ~\eqref{eqn:mfpred}:  ${\kappa = \tilde{\mu_0}/\mu_0 = \mathfrak{F}/\ell^2}$. In order to compute these quantities in our simulation data, we follow the prescription of ~\eqref{eqn:Fcorr}, taking
\begin{equation}
    \frac{\mathfrak{F}}{\ell^2} = \frac{d}{N_c}\sum_{\langle i,j\rangle} \norm{\mathbf{c}_i - \mathbf{c}_j}^2 ,
\end{equation}
where $N_c$ is the total number of contacts, $\langle i,j\rangle$ denotes contacting neighbors, and we approximate $\ell$ as the average distance between contacting particles. These quantities, calculated for both the Gaussian correlated fields $\mathfrak{F}_{\text{GCF}}$ and the wavelike correlated fields $\mathfrak{F}_{\text{WCF}}$, are shown by the grey and black data points in Fig.~\ref{fig:correlationCompare}c, respectively. We also plot the modulus ratio $\kappa$ as a function of correlation length for both GCF(green) and WCF(magenta) simulations.  Although this is a 2d system far from the infinite-dimensional mean field case, the mean field predictions are fairly close to the WCF data, and also capture the general trend of the GCF data.
 
However, the mean-field prediction is not in quantitative agreement so that ${\mathfrak{F}/\ell^2} \neq \kappa$, suggesting that in low dimensions and in particular at smaller $\xi$ the rescaling of the dynamics cannot be reduced solely to the variance of relative local strains, \textit{i.e.}~from the sole characterization of the input field. Nevertheless, a more general prediction of the mean-field theory is that once the mechanical response at one value $\xi$ is known, all others follow. Thus, one may expect that by using AQS as a reference state we can still collapse low-dimensional simulation data using $\kappa$:
\begin{equation}
\sigma \sim  \frac{\tilde{\sigma}}{ \sqrt{\kappa}}, \hspace{0.5cm} \gamma \sim \tilde{\gamma}\sqrt{\kappa}.
\label{eqn:mfscale}
\end{equation}
For individual response curves, the proper value of $\kappa$ is defined at $\tilde\gamma=0$ and computed using the response to small AQS and AQRD strains. Averages can then be taken of these individual curves to obtain bulk behavior. 
Fig.~\ref{fig:correlationCompare}d-f demonstrates that the mean-field prediction works remarkably well: individual stress-strain curves, distributions of strain intervals between stress drops, and the magnitude of stress drops all collapse when properly scaled by $\kappa$ as predicted by mean-field theory. In addition, the collapsed avalanche data is clearly consistent with the scaling of $P(\Delta\sigma) \sim 1/\Delta\sigma$ reported by Shang et. al \cite{shang_elastic_2020}.  This is another indication that bulk responses of AQS and AQRD are controlled by the same physics in that the statistical features of the potential energy landscape are dominated by the scaling of the elastic moduli, \textit{i.e.}~by the curvature of the landscape minima.

\begin{figure}[t]
\centering
\includegraphics[width=0.9\linewidth]{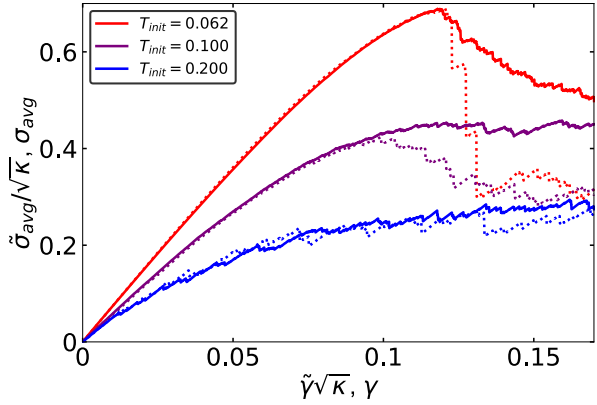}
\caption{\textbf{Average effective stress-strain curves in ultrastable glasses}. Stress-strain curves are shown with $\xi=1$ Gaussian (GCF) driving (solid lines) and AQS driving (dotted lines) on systems which have been prepared via Monte-Carlo Swap at $T_{init}=0.062$ (red), $T_{init}=0.1$ (purple), and $T_{init}=0.2$ (blue). As in Fig.~\ref{fig:correlationCompare}d, $\tilde{\sigma}_{avg}$ and $\sigma_{avg}$ denote an average over configurations but not all elastic branches. The curves are collapsed via ~\ref{eqn:mfscale}. We see that for lower preparation temperatures, there is a larger shear modulus and a more pronounced peak, in accordance with AQS simulations\cite{ozawa_random_2018}. Our predictions for the collapse agree well up to the yielding point ($\gamma \approx 0.12$).}
\label{fig:misakiFigure}
\end{figure}

This is consistent with observations in Fig.~\ref{fig:landscapeStatistics}; the relative offsets between AQRD curves and AQS curves are $\kappa$, $1/\sqrt{\kappa}$, and $\sqrt{\kappa}$ in panels a,b, and c, respectively. Furthermore, this gives additional insight that $\kappa$ remains roughly constant as a function of $N^3p$.

{\em Effect of material preparation and stability---}
To this point, we have investigated infinite-temperature-quenched jammed solids, which have a high degree of disorder. Under AQS, such systems exhibit a ductile yielding transition where the pre-yielding regime transitions smoothly to the post-yielding regime with no discontinuity in the stress. It is well-known that changing the material preparation protocol alters the disorder in the initial configuration, and changes the yielding transition.  Recent work using a new Swap Monte Carlo algorithm generates ultrastable glasses that are -- on the contrary -- extremely brittle, with large stress overshoots and discontinuous stress drops at the yielding transition, and data from such simulations strongly suggests that under AQS the yielding transition is in the Random Field Ising Model universality class~\cite{ozawa_random_2018,popovic_elastoplastic_2018}. Although a full study of the nature of the yielding transition in AQRD is beyond the scope of this work, we analyze the random-stress vs. random-strain curves using GCF under different preparation protocols. 

The solid lines in Fig.~\ref{fig:misakiFigure}a shows such curves for different parent preparation temperatures, ranging from $T_{init} = 0.2$ (ductile glass, low stability) to $0.062$ (brittle, ultrastable glass, high stability). The dashed curves correspond to the stress-strain response in AQS for the same initial conditions.  We observe that in AQRD, the global modulus increases as the stability increases, which is similar to what is observed in AQS. In addition, there is clear stress overshoot (where the average stress increases far above its later steady state value) for the ultrastable glass, which is similar to what is seen for the yielding transition in AQS, although the yielding transition is much sharper in AQS.  Taken together, these results highlight that the qualitative trends for how the yielding transition depends on glass stability are similar in AQS and AQRD, and sets the stage for future work to study the statistics and spatial structure of the yielding transition in AQRD.

\section{Conclusion}
These results demonstrate that shear and random forces perturb disordered solids in remarkably similar ways. In particular, the nonlinear properties of the potential energy landscape traversed by AQS or AQRD display identical scaling exponents. We discovered that the prefactors for these scaling laws, which generally characterize the stiffness of the material or the magnitude of the curvature in the potential energy landscape, are a power law function of the correlation length of the input field of displacements. The exponent ranges from $-1$ to $-3$ depending on the detailed implementation of the field, consistent with the predictions of the mean-field theory. Since AQS corresponds to an input field where the correlation length is the size of the periodic box, it is not special, but instead a terminal point on a family of random fields that can be characterized by their correlation lengths. In general, materials are stiffer in response to fields with smaller correlation lengths.
Conversely, it is more efficient to make a material yield by deforming it in a less correlated way.

Since in the pre-yielding regime AQRD and AQRF generate nearly identical dynamics
-- and AQRF is equivalent to self-propelled particle dynamics in the limit where rotational noise is taken to zero first, and then the self-propelled velocity field is taken to zero -- these results have important implications for the emerging field of dense active matter.  First, it establishes that there is a direct equivalence between sheared and active matter systems in this limit, meaning that decades of work on sheared granular matter can be directly imported to understand active systems. Second, it strongly suggests that the dynamics of dense active matter systems could be predicted using tools already developed for sheared granular systems, such as structural and vibrational mode analyses~\cite{richard_predicting_2020}.  Aspects of such a framework for active matter have already been advanced for instance by Henkes and collaborators~\cite{henkes_active_2011, henkes_dense_2020}. An interesting avenue for future research will be to study how small but finite particle velocities introduce fluctuations into the system that perturb this equivalence to shear. Does this create features analogous to those in finite-strain rate shear simulations? Additionally, we could introduce small but finite magnitudes of rotational noise so that the input displacement fields rotate over time, instead of remaining fixed indefinitely as presented here. We speculate that such dynamics could also be very similar to sheared systems at finite strain rates and/or in the presence of perturbative thermal noise, another active area of research in the rheology community. In experiments, it should be possible to quantify the random stress we define here by studying active photoelastic disks where the internal stress in the system can be inferred from light patterns.

A second obvious avenue for future work is to understand the spatial structure and the nature of the yielding transition under AQRD.  Our work confirms that the basic phenomenology is the same: there is a yielding transition where the macroscopic rheology of the material switches from elastic (stress proportional to strain) to fluid-like (stress independent of strain), the macroscopic modulus of the material before it yields is different from the local modulus along elastic branches, and the nature of the yielding transition changes as a function of material preparation. However, this opens more questions than it answers, such as: what are the correlation lengths of the \emph{output} particle displacement fields that occur in response to the input displacement fields we study? An emerging body of work has begun to show that such correlations tend to long-range and depend on the distance to an instability~\cite{bi_motility-driven_2016,henkes_dense_2020,szamel_long-ranged_2021-1}, making any relationship to the input field non-trivial. Is the yielding transition under AQRD still in the Random Field Ising Model universality class? Under AQS, brittle glasses fail via localized shear band where all the strain is accommodated in a small region of the material -- is something similar true in AQRD? Do we have to re-define "localized" to account for the fact that there is no macroscopic symmetry for AQRD with random Gaussian input fields? Does localization depend on the correlation length of the input field? Such questions are more than academic, as they help us to predict how dense materials composed of active matter flow and fail.  Answering them will help us to harness the activity of active matter to develop actuated solids that can perform tasks, or even predict emergent collective phenomena in crowded active matter systems.

\matmethods{Simulations were performed using pyCudaPack (https://github.com/SimonsGlass/pyCudaPacking/) and monteCarloPack (https://github.com/SimonsGlass/monteCarloPCP/) which are available upon request.}


\showmatmethods{} 

\acknow{We thank Ludovic Berthier and Misaki Ozawa for discussion and initial configurations of ultrastable glasses using swap monte carlo. PKM would like to thank Matthias Merkel and Brian Tighe for helpful discussions.
EA would like to thank Francesco Zamponi and Ada Altieri for discussions about the infinite-dimension mean-field results. EA acknowledges support from the Swiss National Science Foundation by the SNSF Ambizione Grant PZ00P2{\_}173962,
from the European Research Council (ERC) under the European Union Horizon 2020 research and innovation programme (grant agreement n.~723955 - GlassUniversality),
and from the Simons Foundation Grant (\#454955 Francesco Zamponi).
EC acknowledges support from the Simons Foundation Grant \#454939, MLM, PKM and SR acknowledge support from from Simons Foundation Grants \#46222 and \#454947, and MLM acknowledges support from NSF-DMR-1951921.}

\showacknow{} 

\bibliography{main}

\end{document}




\SItext

\setcounter{equation}{0}
\renewcommand{\theequation}{S\arabic{equation}}%

\section{Generating random vectors for AQRD}

\subsection{Wave-like correlated fields (WCF)}\label{sec_rf}

We generate the vector $\ket{c}$ under the constraint that it must be continuous across the boundary, and thus periodic. In order to generalize AQS strain, we set the $y$-component of each $\mathbf{c}_i$ to zero and let the $x$-component depend on the height of the particle 
\begin{equation}
    \mathbf{c}_i = \sin{\bigg( \frac{\pi y_i}{\xi}\bigg)} \hat{x}
\end{equation}
where $\hat{x}$ is the unit vector in the $x$-direction and $\xi$ is the correlation length. In order for this to be periodic, we must have $\xi = L_y/n$ with $n \in \mathbb{Z}$. We note that AQS with Lees-Edwards boundary conditions is simply $\xi = L_y$ with a phase shift (or simply a cosine). Once $\ket{c}$ is determined, it is normalized so that $\braket{c}{c}=1$.

\subsection{Gaussian correlated fields (GCF)}\label{FT}
This section describes the process of generating a random Gaussian vector $\ket{c}$ with finite spatial correlation length~$\xi$ using Fourier transforms. This method is used for all GCF ensembles, except $\xi = 1$ (corresponding to completely uncorrelated field) where each component of $\ket{c}$ is drawn from a uniform distribution and then normalized to length $1$.

The two-dimensional system is a box of size ${L_x \times L_y}$ with periodic boundary conditions, allowing us to define wave vectors $\mathbf{k}_{nm} = \big(\frac{a \pi n}{L_x},\frac{b \pi m}{L_y}\big)$ for $n,m \in \mathbb{Z}$ where $a=\pm 2$ and $b=\pm 2$, with signs decided randomly, as discussed below. In practice, we truncate the Fourier sums at ${n,m = Q}$ taking ${Q=20}$.

To create $\ket{c}$, we need a correlated random field ${\Psi(\mathbf{x})}$ which is Gaussian distributed with zero mean ${\langle\Psi(\mathbf{x}) \rangle = 0}$ and has the two-point correlator $\langle \Psi({\bf x})\Psi({\bf x}')\rangle = f(\vert {\bf x-x}'\vert)$.
We enforce ${f(x)}$ to be a Gaussian function,  
whose explicit Fourier transform is
${\tilde{f}(|{\bf k}|) = \exp[(-\frac{{| {\bf k}|}^2\xi^2}{(a^2 + b^2)})]}$. 

First, we generate a set of uncorrelated random fields ${\tilde{\psi}({\bf k})}$
with $\langle \tilde{\psi}({\bf k})\rangle = 0$
and 
$\langle \tilde{\psi}({\bf k})\psi({\bf k}')\rangle = \frac{1}{4 \pi^2} \delta_{{\bf k} , -{\bf k}'}$, 
where 
$\delta$ is the Kronecker function, and the factor ${4 \pi^2}$ comes from the Fourier transform convention.
In practice, for each wave vector of the truncated sum we generate a random field ${\tilde{\psi} (\mathbf{k}_{nm}) = A({\bf k}) \exp{(iB(\bf{k}))}}$,
with ${A({\bf k})=A(-{\bf k})}$ normally distributed with zero mean and variance
$\frac{1}{4\pi^2}$,
and ${B({\bf k}) = -B(-{\bf k})}$ uniformly distributed on the interval ${[0,2\pi]}$.

Secondly, we use the Fourier transform of the target correlator ${f(x)}$ to construct a new field
${\tilde\Psi({\bf k}) = \tilde{f}(|{\bf k}|) \, \tilde \psi({\bf k})}$,
whose Fourier transform of the $\alpha$ component is
\begin{equation}
\begin{split}
	\Psi^\alpha(\mathbf{x})
		= \sum^Q_{n,m = 1} A_{nm}^\alpha & e^{-{|\mathbf{k}_{nm}|}^2\xi^2/(a^2+b^2)}\cos{(B_{nm}^\alpha+ \mathbf{k}_{nm} \cdot \mathbf{x})}.
\end{split}
\end{equation}
The random vectors $\mathbf{c}_i = \big(c_i^x,c_i^y\big)$ are then defined as $c_i^\alpha = \Psi^\alpha(\mathbf{x}_i)$ using the initial positions $\mathbf{x}_i$. Once $\ket{c}$ is determined, it is normalized so that $\braket{c}{c}=1$.

It is important to note that because these fields are built using Fourier transforms, they will have a bulk phase preference at high values of $\xi$ -- greater than about a quarter of the box size -- along the $\tan^{-1}\big(\frac{b}{a}\big)$ axis. In other words, the mulitiplicity of fields that obey the necessary conditions -- that they have zero mean and they respect the periodic boundary conditions -- becomes very small as $\xi$ approaches the box size, and in fact in the limit that $\xi$ equals the box size, only the simple shear strain field satisfies those conditions.

Therefore, for GCF fields we restrict ourselves to values of $\xi$ for which this bulk phase preference does not dominate. It is also for this reason that the signs of $a$ and $b$ are chosen at random for each instantiation of a system. Other even integer values of $a$ and $b$ may be used to create Gaussian correlated fields, but these fundamentally alter the symmetry of the system for high values of $\xi$, and thus are not used in this work.

\section{Athermal quasistatic random forcing (AQRF)}
In addition to the strain-controlled definition of AQRD in the main text, we can also consider a stress-controlled version denoted athermal quasistatic random forcing (AQRF). Instead of enforcing a displacement vector $\ket{c}$, we apply an external force $\ket{F^{ext}} = f \ket{c}$ on the system and we measure the resulting strain, where again $\braket{c}{c}=1$. Thus, instead of a constrained minimization, we simply perform a minimization subject to a fixed external force. Once minimized, the sum of all forces on each particle must be zero:
\begin{equation}
    \mathbf{F}_i = \sum_{j \in \partial i} \mathbf{F}_{ij} + f \mathbf{c}_i = 0,
    \label{eqn:forceSum}
\end{equation}
where $j \in \partial i$ indicates a sum over interparticle forces acting on particle $i$. The sum of interparticle forces is parallel to $\ket{c}$, allowing closure with $\ket{c}$ to give $f = -\braket{c}{F}$. Following the arguments around Eq.~(2) a normalization factor is necessary to compare the size of stresses in AQS leading to
\begin{equation}
\tilde\sigma = -\frac{f}{L_x} \sqrt{\frac{N}{12}}.    
\end{equation}
The random strain is then naturally defined as %
\begin{equation}
\tilde\gamma = \frac{1}{A}\frac{d U}{d \tilde\sigma} = \frac{1}{L_xL_y}\sum_{i=1}^N \frac{\partial U}{\partial \mathbf{x}_i}\cdot\frac{d\mathbf{x}_i}{d\tilde\sigma}.
\end{equation}
where ~\eqref{eqn:forceSum} ensures that ${\frac{\partial U}{\partial \mathbf{x}_i} = \sum_{j \in \partial i} \mathbf{F}_{ij} = -f\mathbf{c_i}}$, and $\frac{d\mathbf{x}_i}{d\tilde\sigma}$ can be interpreted for small stresses as the displacement $\Delta \mathbf{x}_i$ of particle $i$, subject to the applied stress $\tilde\sigma$, leading to the natural relation
\begin{equation}
    \frac{d\mathbf{x}_i}{d\tilde\sigma} = - \frac{\Delta \mathbf{x}_iL_x}{f\sqrt{\frac{N}{12}}}.
\end{equation}
Together these lead to the definition of random strain in the stress-controlled ensemble:
\begin{equation}
    \tilde\gamma = \frac{\bra{c}\ket{\Delta x}}{L_y\sqrt{\frac{N}{12}}}
\end{equation}
Physically, this scalar product is simply the sum over all particles of their individual displacements along their corresponding imposed direction $\mathbf{c}_i$. This is the natural counterpart of the random stress definition given in Eq.~(5).


\section{Generality of the stress-strain collapse}

\begin{figure}[htp]
\centering
\includegraphics[width=0.9\textwidth]{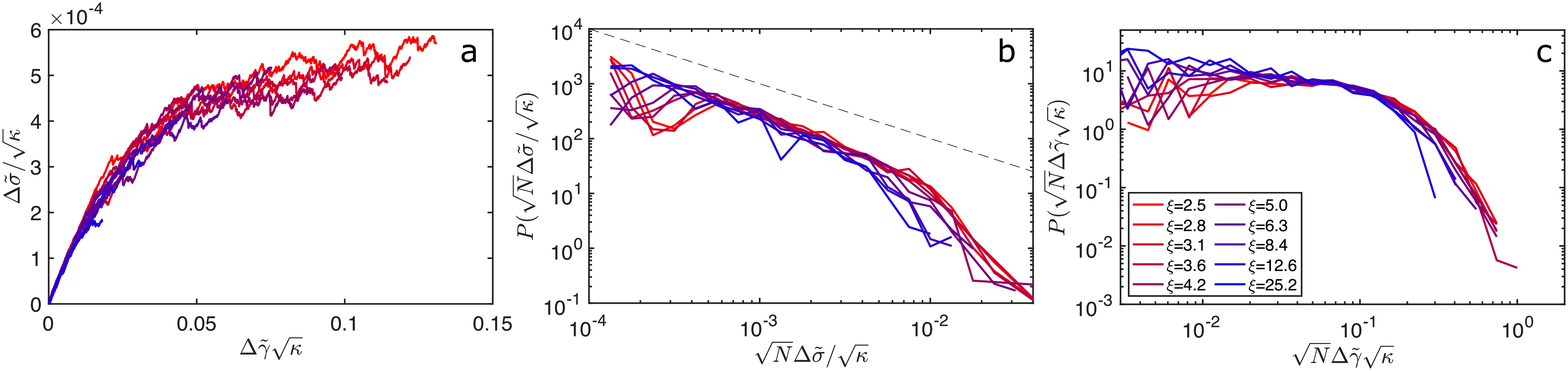}
\caption{The same scaling which in Fig. 5 collapses the AQRD data for Gaussian Correlated Fields is applied to Wave-like Correlated Fields (WCF) of varying correlation length $\xi$. Here the \textbf{a)}~average stress-strain curves, \textbf{b)}~distribution of the stress drops, and \textbf{c)}~the distribution of strain intervals between events all collapse when scaled with the ratio of initial shear moduli $\tilde{\mu}_0$ vs. the AQS initial shear modulus $\mu_0$ as $\kappa = \tilde{\mu}_0/\mu_0$. The dotted line in panel b is of slope $-1$. All systems have $N=2048$ particles.}
\label{fig:wcf}
\end{figure}

\begin{figure}[htp]
\centering
\includegraphics[width=0.9\textwidth]{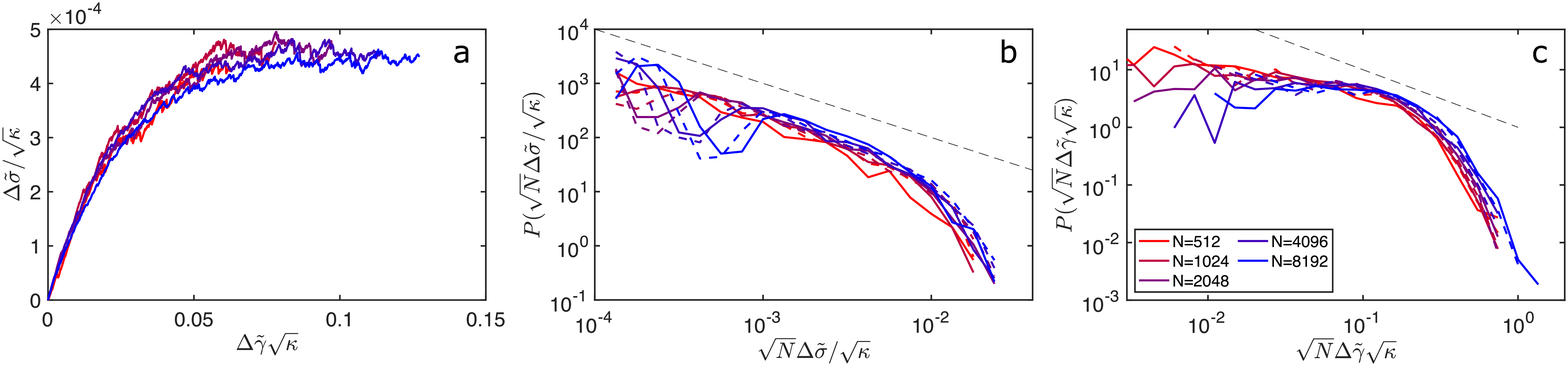}
\caption{The same scaling which in Fig. 5 collapses the AQRD data for Gaussian Correlated Fields with $N=2048$ particles is applied to systems of varying size. Here the \textbf{a)}~average stress-strain curves, \textbf{b)}~distribution of the stress drops, and \textbf{c)}~the distribution of strain intervals between events all collapse when scaled with the ratio of initial shear moduli $\tilde{\mu}_0$ vs. the AQS initial shear modulus $\mu_0$ as $\kappa = \tilde{\mu}_0/\mu_0$. Dashed lines in b and c represent AQS systems of the same size. The dotted line in panel b is of slope $-1$. All systems use GCF fields with $\xi=1$. While curves are generated with the same protocol detailed in the text, it is clear that the dynamic range of the strain should be applied to a maximum value which is proportional to $\sqrt{N}$.}
\label{fig:finSize}
\end{figure}

In Fig. 5d-f, we show the collapse of the average stress-strain curves, the probability distribution of stress drops, and the probability distribution of the strain intervals between rearrangements. Here, we show those same curves for both WCF fields (Fig.~\ref{fig:wcf}) and for GCF systems of different size at constant correlation length (Fig.~\ref{fig:finSize}). Empirically, Fig.~\ref{fig:finSize}b shows that the distribution of stress drops and the distribution of strain intervals between rearrangements collapse when scaled as ${\sqrt{N}\Delta\tilde{\sigma}/\sqrt{\kappa}}$ and ${\sqrt{N}\tilde{\gamma}\sqrt{\kappa}}$, in contrast to the geometric mean values $\langle\tilde{\sigma}\rangle \sim p/N$ and $\langle\tilde{\gamma}\rangle \sim p^{1/3}/N$. This seeming discrepancy is due to the choice of a fixed strain step size which systematically under-counts any small rearrangements. The full distribution appears to be a power law distribution with a minimum stress drop known to scale as $1/N$ (see supplement of Ref.~\cite{morse_differences_2020}). Thus, instead of fitting the center of the distribution, we ignore the small cutoff and match the high tail. 

\section{Scaling argument supporting our mean-field prediction in finite dimension}

In order to gain more physical intuition on our infinite-dimensional mean-field results,
we present thereafter a scaling argument at a finite dimension $d$ which supports the mean-field prediction that the total effective strain would be given by
${\gamma_{\text{eff}}=\tilde{\gamma} \, \sqrt{\mathfrak{F}}/\ell}$ and that the typical elastic modulus should scale as ${\mu \sim \mathfrak{F}/\ell^2}$.
The different assumptions underlying this argument turn out to be exact in infinite-dimension, where we were able to obtain this prediction from the computations, even before having any physical  intuition of what to expect.

Let's first consider the most extreme case with an infinite $\xi$. This means that all particles are driven with a same vector ${\bf c}_i$: the relative local strains are all strictly zero, with no variance (${\mathfrak{F}=0}$)
or equivalently a distribution of relative strains ${\bar{\mathcal{P}}({\bf c}_{ij})=\delta({\bf c}_{ij})}$.
The whole system is thus simply translated in space, and its effective strain always remains strictly zero.

\begin{figure}[h]
\begin{center}
\includegraphics[width=0.7\textwidth]{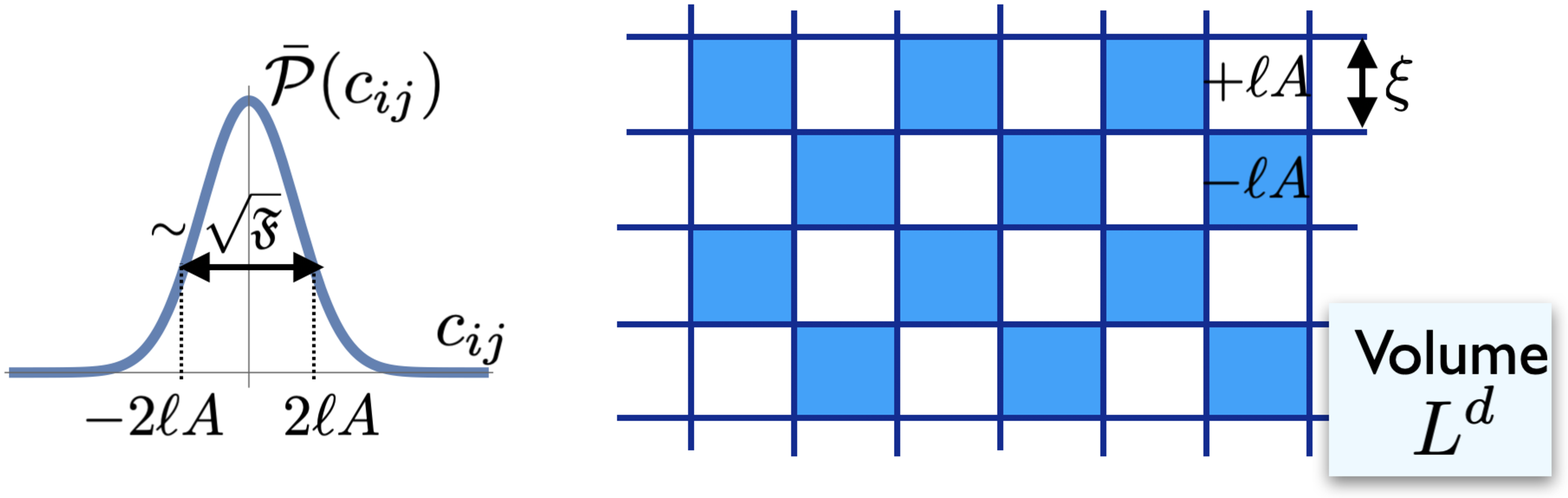}
\end{center}
\caption{
\textit{Left:}~If we have a Gaussian distribution of relative strains of zero mean and variance $\mathfrak{F}$,
we can simplify it in a patchy picture where we retain only the value at the peak (${{\bf c}_{ij}=0}$) or its standard deviation (${{\bf c}_{ij} = \pm 2 \ell A }$).
\textit{Right:}~The associated patchy representation of the displacement field itself, where ${{\bf c}_{i} = \pm \ell A}$ on alternating patches.
}
\label{fig:scaling-MF}
\end{figure}

If instead we allow for a large but finite $\xi$,
the system can be pictured as being composed of large `patches' in which particles share the same vector ${\bf c}_i$, and pairs have thus zero relative strains ${{\bf c}_{ij}}$ inside a given patch.
Only pairs of particles living at the boundaries between such patches experience a non-zero ${{\bf c}_{ij}}$, and can thus contribute to the total effective strain felt by the system under AQRD.
The smaller $\xi$,
the larger the proportion of pairs at a boundary.
We can quantify this within the simplified patchy picture:
let's assume that we alternate in 2d, as in a chessboard, patches with ${{\bf c}_i=\pm {\bf A}}$.
Inside a patch ${{\bf c}_{ij}=0}$;
pairs at the boundaries have ${{\bf c}_{ij}=\pm 2{\bf A}}$,
as illustrated in Fig.~\ref{fig:scaling-MF}.
We can estimate the proportion of interacting pairs living at a boundary as
${\rho = L/\xi \times \ell \times L \times 2/L^2 = 2 \ell/\xi}$.
In dimension $d$ this generalizes to ${\rho=d \, \ell/\xi}$.
The variance of relative strains is then given by
$$ \mathfrak{F}/\ell^2
= \sum_{\text{interacting pairs}} {\bf c}_{ij}^2 
= \sum_{\text{in a patch}} (1-\rho) \times 0^2
    + \sum_{\text{in a patch}} \rho \times (2 {\bf A})^2
= 4  A^2 d \ell/\xi \, .
$$
Consequently, because the relative strain on interacting pairs is \underline{distributed},
the typical strain applied to a given pair scales as
${\sim \tilde{\gamma} \, \sqrt{\mathfrak{F}}/\ell
= \tilde{\gamma} \, 2 A \sqrt{d \ell/\xi}}$.

Beyond this simplified patchy picture,
Eq.~(6) in our manuscript gives the \underline{exact} variance ${\mathfrak{F}}$ for a spatially-correlated input field.
If we assume that relative strains have a Gaussian distribution ${\bar{\mathcal{P}}(c_{ij})}$ of zero mean and variance $\mathfrak{F}$, 
the $A$ in our patchy picture corresponds to the standard deviation ${\sqrt{\mathfrak{F}}}$, as illustrated in Fig.~\ref{fig:scaling-MF}.
Assuming that the correlator $f_\xi(x)$ is a Gaussian function (as in our Eq.~(5)):
if ${\xi/\ell \ll 1}$, $A$ is essentially a constant;
if ${\xi/\ell \gg 1}$, we have instead $A \sim \ell/\xi$.
The latter case adds an additional dependence on $\xi$ on the variance $\mathfrak{F}$, and thus on the typical strain felts by an interacting pair.
That way we recover the crossover from ${\mathfrak{F} \sim \ell/\xi}$ to ${\mathfrak{F} \sim (\ell/\xi)^3}$ that we discuss in our manuscript.

In lower dimensions, these assumptions can be assumed to hold at least in the pre-yielding regime that we consider, on each elastic branch (albeit if we start to interfere with the spatial correlations of the response field after an AQRD step, as mentioned in our conclusion).
In AQRD, after each minimization step, 
the random stress is given by our Eq.~(4), as the scalar product of the forces acting on each particles ${\bf F}_i$ and their respective vector ${\bf c}_i$. This can be rewritten as a scalar product of the forces between pairs and their respective relative strain ${\bf c}_{ij}$:
$$
\sum_{\langle ij \rangle} {\bf F}_{ij} \cdot {\bf c}_{ij}
= \frac12 \sum_{i,j=1}^N {\bf F}_{ij} \cdot \left( {\bf c}_i - {\bf c}_j \right)
= \frac12 \sum_i \underbrace{\left( \sum_j {\bf F}_{ij} \right)}_{={\bf F}_i} \cdot {\bf c}_i
+ \frac12 \sum_j \underbrace{\left( \sum_i {\bf F}_{ji} \right)}_{={\bf F}_j} \cdot {\bf c}_j
=\sum_{i} {\bf F}_i \cdot {\bf c}_i
\propto - \tilde{\sigma}
$$
where in the last equality we skipped the normalization with respect to the system size and the number of particles (see Eq.(4)).
Let's assume that we start from a minimum after an AQRD step, described by a given set of relative position $\lbrace {\bf r}_{ij} \rbrace$.
If we apply an infinitesimal strain increment ${\Delta \tilde{\gamma} }$ that keeps us in an elastic branch, we can Taylor-expand the forces using
$$
\nabla v (\vert {\bf r}_{ij} + \Delta \tilde{\gamma} \, {\bf c}_{ij} \vert)
= \nabla v (\vert {\bf r}_{ij} \vert)
    + \Delta \tilde{\gamma} \, \left[
        v''(r_{ij}) \, (\hat{\bf r}_{ij} \cdot {\bf c}_{ij}) \, \hat{\bf r}_{ij}
        + v'(r_{ij}) \, \frac{{\bf c}_{ij}}{r_{ij}}
    \right] + \mathcal{O} \left(\Delta \tilde{\gamma}^2 \right) \, .
$$
The associated increase in random stress is then given by (we skip the finite-size normalization thereafter): 
\begin{equation}
\nonumber
\begin{split}
\Delta \tilde{\sigma}
&\propto - \sum_{\langle ij \rangle} \left[{\bf F}_{ij}(\Delta \tilde{\gamma}) - {\bf F}_{ij}(\Delta \tilde{\gamma}=0) \right] \cdot {\bf c}_{ij}
\\
&= \Delta \tilde{\gamma} \, \sum_{\langle ij \rangle} \left[
    v''(r_{ij}) \, (\hat{\bf r}_{ij} \cdot {\bf c}_{ij}) \, \hat{\bf r}_{ij}
        + v'(r_{ij}) \, \frac{{\bf c}_{ij}}{r_{ij}}
\right] \cdot {\bf c}_{ij}
= \Delta \tilde{\gamma} \, \underbrace{\sum_{\langle ij \rangle} c_{ij}^2 \left[
    v''(r_{ij}) \, (\hat{\bf r}_{ij} \cdot \hat{{\bf c}}_{ij})^2
        + \frac{v'(r_{ij})}{r_{ij}}
\right]}_{\text{local elastic modulus}}
\, .
\end{split}
\end{equation}
This local elastic modulus depends on the distribution of ${\lbrace {\bf r}_{ij} \rbrace}$, which is a highly nontrivial quantity to characterize analytically.
In fact, the infinite-dimensional limit is the only case where we know it exactly, as a function of accumulated strain, which is one of the reasons why it is such a precious benchmark.
Finally, if we assume that the typical local elastic modulus scales as the average $c_{ij}^2$, by definition this is equal to the variance $\mathfrak{F}/\ell^2$.
So we recover that the typical elastic modulus should scale as ${\mu \sim \mathfrak{F}/\ell^2}$, and thus inherits the $\xi$-dependence of ${\mathfrak{F}}$ along the way.
Note that we did not use any property specific to the Hertzian interaction potential, our argument holds for a generic soft potential.





\bibliography{supplement}